\documentclass[fleqn,twoside]{article}
\usepackage[headings]{espcrc2}

\readRCS
$Id: espcrc2.tex,v 1.2 2004/02/24 11:22:11 spepping Exp $
\usepackage{graphicx}
\usepackage[figuresright]{rotating}

\newcommand{\AmS}{{\protect\the\textfont2
  A\kern-.1667em\lower.5ex\hbox{M}\kern-.125emS}}

\title{Bottonium mass -- evaluation using renormalon cancellation}

\author{C. Contreras\address[MCSD]{Physics Department, Universidad T\'ecnica Federico Santa Mar\'{\i}a, Valpara\'{\i}so, Chile},
        G. Cveti\v c\addressmark\thanks{The work supported in 
part by FONDECYT grant No.~1010094 (G.C.), and project 
USM No.~110321 of the UTFSM (C.C. and G.C). Talk presented by G.C., at QCD04, Montpellier, France, July 5-9, 2004.},
        and
        P. Gaete\addressmark[MCSD]}
       
\begin{document}

\begin{abstract}
We present a method of calculating the bottonium mass $M_{\Upsilon}(1S)
= 2 m_b\!+\!E_{b \bar b}$. The binding energy is separated into the soft 
and ultrasoft components $E_{b \bar b}=E_{b \bar b}(s)+E_{b \bar b}(us)$
by requiring the reproduction of the correct residue parameter value
of the renormalon singularity for the renormalon cancellation
in the sum $2 m_b\!+\!E_{b \bar b}(s)$.
The Borel resummation is then performed separately for
$2 m_b$ and $E_{b \bar b}(s)$, using the infrared safe
${\overline m}_b$ mass as input. $E_{b \bar b}(us)$ is estimated.
Comparing the result with the measured value of $M_{\Upsilon}(1S)$,
the extracted value of the quark mass is  
${\overline m}_b(\mu\!=\!{\overline m}_b) = 4.241 \pm 0.068$ GeV
(for the central value $\alpha_s(M_Z)=0.1180$).
This value of ${\overline m}_b$ is close to the earlier values 
obtained from the QCD spectral sum rules, but lower than from pQCD 
evaluations without the renormalon structure for heavy quarkonia.
\vspace{1pc}
\end{abstract}

\maketitle


Heavy quarkonia $q \bar q$ ($q=b, t$) can be investigated
by perturbative methods (pQCD) via effective theories
NPQCD \cite{Caswell:1985ui} and pNRQCD \cite{Pineda:1997bj}
(or: vNRQCD \cite{Luke:1999kz}) because of the scale
hierarchies of the problem:
$m_q > m_q\alpha_s(\mu_s) > m_q \alpha_s^2(\mu_{us}) \stackrel{>}{\sim}
\Lambda_{\rm QCD}$. Here, $m_q$ is the (pole) mass of the
quark, $\mu_s\!\sim\!m_q\alpha_s(\mu_s)$ is the
soft, and $\mu_{us}\!\sim\!m_q \alpha_s^2(\mu_{us})$ the
ultrasoft energy. The quarkonium mass is $M_{q \bar q} = 2 m_q +
E_{q \bar q}$, where the binding energy consists of the soft and
ultrasoft regime contributions: $E_{q \bar q} = E_{q \bar q}(s) +
E_{q \bar q}(us)$. A practical problem which appears in the
course of evaluation of $M_{q \bar q}$ is that the perturbative pole 
mass has an inherent ambiguity $\delta m_q \sim \Lambda_{\rm QCD}$
($\sim\!0.1$ GeV) due to the infrared (IR) renormalon singularity
which appears at the value of the Borel transform variable $b\!=\!1/2$
for $m_q/{\overline m}_q$. Here, ${\overline m}_q$ is the infrared safe
(renormalon-free) ${\overline {\rm MS}}$ mass. However, the static
potential $V_{q \bar q}(r)$ has a related ambiguity
$\delta V_{q \bar q}(r)\!\sim\!\Lambda_{\rm QCD}$ such that
$\delta(2 m_q + V_{q \bar q}) = 0$, i.e., the
renormalon singularity cancels for the combined quantity
$2 m_q\!+\!V_{q \bar q}$ \cite{Hoang:1998nz}
(see also \cite{Neubert:1994wq}). The static potential
is a quantity which does not contain ultrasoft regime
contributions \cite{Brambilla:1999qa}. Therefore, $E_{q \bar q}(s)$
contains the entire $V_{q \bar q}$ and kinetic
energy effects, the latter are renormalon-free. Thus, the $b\!=\!1/2$
renormalon singularity of $V_{q \bar q}$ and $E_{q \bar q}(s)$
are equal and hence the singularity must
cancel also in the combination $2 m_q\!+\!E_{\rm q \bar q}(s)$
\begin{equation}
\delta\left[ 2 m_q + E_{\rm q \bar q}(s) \right] = 0 \ .
\label{rencanc}
\end{equation}  
In principle, $E_{q \bar q}(us)$ could be included in
this relation. However, in practice, it distorts the
cancellation effects since we know these quantities only
to a finite order in perturbation expansions
[cf.~the discussion following Eq.~(\ref{kappa})].

One possibility of evaluating the bottonium ground state
$\Upsilon(1S)$ mass is to use an infrared-safe (renormalon-free)
quark mass (${\overline m}_b$, $m_b^{\rm RS}$, etc.) and
a common couplant $a(\mu) = \alpha_s(\mu)/\pi$ as inputs
in the evaluation of the available truncated perturbation expansion
(TPS) for $M_{\Upsilon}(1S) = 2 m_b + E_{b \bar b}$, in order
to avoid the $b\!=\!1/2$ renormalon (divergence) problems throughout
\cite{Pineda:2001zq}, and then extract the value of ${\overline m}_b$
from the measured value $M_{\Upsilon}(1S) = 9460$ MeV.

Another possibility is to evaluate $E_{b \bar b}$ in terms of the
pole mass $m_b$ and of $a(\mu)$, taking into account the
$b=1/2$ singularity of $E_{b \bar b}$ (using, e.g., the Principal Value
[PV] prescription in the Borel integration), and adding $2 m_b$ to
$E_{b \bar b}$. From $M_{\Upsilon}(1S) = 
2 m_b^{\rm (PV)}\!+\!E_{b \bar b}(m_b^{({\rm PV})})$,
the PV-value of the pole mass $m_b$ is then extracted, and subsequently
the value of ${\overline m}_b$ (via PV Borel integration prescription).
This is the approach of Ref.~\cite{Lee:2003hh}.

Our approach \cite{CCG} follows to a significant degree the 
latter approach, but with some important modifications:
\begin{enumerate}
\item
The input parameter is the renormalon-free mass
${\overline m}_b \equiv {\overline m}_b(\mu\!=\!{\overline m}_b)$ 
(and, of course, the QCD couplant $a(\mu)$).
The pole mass $m_b=m_b({\overline m}_b;\mu_m)$ is evaluated via Borel
integration, accounting for the $b=1/2$ singularity,
and using a hard renormalization scale $\mu_m\!\sim\!m_b$.
The residue parameter $N_m$ of the $b\!=\!1/2$ singularity is
evaluated from the available TPS for $m_b/{\overline m}_b$.
\item
On the basis of the knowledge of $N_m$, we separate
the binding energy into the soft ($s$) and ultrasoft ($us$) regime
contributions: $E_{b \bar b} = 
E_{b \bar b}(s; \mu_f)\!+\!E_{b \bar b}(us; \mu_f)$, 
where the $s$-$us$ factorization
scale $\mu_f$ parametrizes the separation. The separation is
performed by accounting for the renormalon cancellation in the sum 
$2 m_b\!+\!E_{b \bar b}(s)$: $N_m(m_b)\!=\!N_m(E_{b \bar b}(s;\mu_f))$.
The latter relation fixes $\mu_f$ and thus the separation.
\item
The soft binding energy $E_{b \bar b}(s;\mu_f)$ is then
evaluated via Borel integration, accounting for the $b=1/2$
singularity (using the same prescription, e.g. PV, as for $m_b$),
and using a soft renormalization scale $\mu_s\!\sim\!m_b\alpha_s$.
\item
The value of the ultrasoft part $E_{b \bar b}(us; \mu_f)$ 
is estimated.
\item
{}From $2 m_b + E_{b \bar b}(s) + E_{b \bar b}(us) = M_{\Upsilon}(1S)$,
the value of ${\overline m}_b$ is extracted.
\end{enumerate}
For details, we refer to Ref.~\cite{CCG}.

\section{Evaluation of $m_b$ and $N_m$}
\label{sec:mb}

This part has been performed mostly in 
Refs.~\cite{Pineda:2001zq,Lee:2003hh,Cvetic:2003wk}.
The pole mass is known to NLO:
\begin{equation}
S \equiv \frac{m_b}{{\overline m}_b} - 1 = \frac{4}{3}a(\mu_m)
\sum_{j=0}^{\infty} a^j(\mu_m) r_j(\mu_m) \ ,
\label{mbexp}
\end{equation}
where $r_1$ and $r_2$ are known coefficients ($r_0=1$), e.g., in the
${\overline {\rm MS}}$ scheme, and they
depend on $(\mu_m/{\overline m}_b)$; $\mu_m\!\sim\!{\overline m}_b$. 
The Borel transform is
\begin{eqnarray}
\lefteqn{
B_S(b) = \frac{4}{3} \left[ 1 + \frac{r_1}{1! \beta_0} b +
\frac{r_2}{2! \beta_0^2} b^2 + {\cal O}(b^3) \right] }
\label{Bmb1}
\\ 
& = & \frac{ N_m \pi \mu_m}{{\overline m}_b (1\!-\!2 b)^{1\!+\!\nu}}\!
\sum_{k=0}^{\infty} {\tilde c}_k (1\!-\!2 b)^k\!+\!B_S^{\rm (an.)}(b) ,
\label{Bmb2}
\end{eqnarray}
where $\beta_0\!=\!(11\! -\! 2 n_f/3)/4$, 
$\beta_1\! =\! (102\! -\! 38 n_f/3)/16$, 
$\nu\!=\!\beta_1/(2 \beta_0^2)$ ($n_f\!=\!4$); 
${\tilde c}_0\!=\!1$ and the next three coefficients ${\tilde c}_k$
are known (\cite{Pineda:2001zq} for $k\!=\!1,2$; \cite{CCG} for $k\!=\!3$).
$B_S^{\rm (an.)}(b)$ is the analytic part in the bilocal
expansion (\ref{Bmb2}) \cite{Lee:2003hh}, and it is
known up to $\sim\!b^2$. The residue parameter $N_m$ in Eq.~(\ref{Bmb2})
can be obtained with high precision 
\cite{Pineda:2001zq,Lee:2003hh,Cvetic:2003wk}
\begin{equation}
N_m = \frac{{\overline m}_q}{\mu_m} \frac{1}{\pi} R_S(b=1/2) \ ,
\label{Nmform}
\end{equation}
where, according to (\ref{Bmb2})
\begin{equation}
R_S(b; \mu_m) \equiv  (1 - 2 b)^{1 + \nu} B_{S}(b; \mu_m) \ .
\label{RSm}
\end{equation}
Applying the Pad\'e P[1/1] to the known NNLO TPS of
$R_S(b)$ then gives
\begin{equation}
N_m(n_f\!=\!4) = 0.555 \pm 0.020 \ .
\label{Nmnf4}
\end{equation}
The pole mass $m_b$, with ${\overline m}_b$ and $a(\mu_m)$
as input, can now be evaluated by Borel integration using
the bilocal expression (\ref{Bmb2})
\begin{equation}
S(b) = \frac{1}{\beta}_0 {\rm Re}\! 
\int\! db 
\exp \left( - \frac{b}{\beta_0 a(\mu_m)} \right) B_S(b; \mu_m) \ ,
\label{BSint}
\end{equation}
where the integration path can be taken along a ray in the
first or fourth quadrant 
(the generalized PV prescription 
\cite{Caprini:1998wg,Cvetic:2001sn,Cvetic:2002qf}).

\section{Separation} 
\label{sec:sep}

The TPS of the binding energy $E_{b \bar b}$
\begin{equation}
E_{b \bar b} = - \frac{4 \pi^2}{9} {\overline m}_b a^2 \sum_{k=0}^{\infty}
a^k f_k \ ,
\label{Ebbexp}
\end{equation}
is known to the impressive order ${\cal O}(m_b a^5)$
\cite{Gupta:pd,Titard:1993nn,Czarnecki:1997vz,Melnikov:1998ug,Penin:1998zh,Kniehl:2002br,Penin:2002zv},
i.e., in Eq.~(\ref{Ebbexp}) $f_k$ ($k=1,2,3$) are known ($f_0=1$). 
The renormalization scale used in
expansion (\ref{Ebbexp}) should be soft ($\mu_s\!\sim\!m_b\alpha_s$)
or lower. The ultrasoft contributions appear for the first
time at $\sim\!m_b a^5$ \cite{Kniehl:2002br,Penin:2002zv}, 
i.e., $f_3\!=\!f_3(s)\!+\!f_3(us)$. The
$us$ coefficient can be written as \cite{CCG}: 
$f_3(us)/\pi^3 =27.5\!+\!7.1 \ln \alpha_s(\mu_s)\! -\!14.2 \ln \kappa$,
where $\kappa\!\sim\!1$ is the parameter of the $s$-$us$
factorization scale $\mu_f$: $\mu_f\! =\!\kappa m_b \alpha_s(\mu_s)^{3/2}$.
It can be fixed by the requirement of the renormalon
cancellation in $2 m_b\!+\!E_{b \bar b}(s)$:
\begin{equation}
N_m = \frac{2 \pi}{9} \frac{ {\overline m}_b a(\mu_s) }{\mu_s}
R_{F(s)}(b;\mu_s;\mu_f) {\big |}_{b=1/2} \ ,
\label{NmEqq}
\end{equation}
where, in analogy with $R_S$ of (\ref{RSm}) 
\begin{equation}
R_{F(s)}(b; \mu_s; \mu_f) = (1 - 2 b)^{1 + \nu} B_{F(s)}(b;\mu_s; \mu_f) \ ,
\label{RFs}
\end{equation}
and $B_{F(s)}$ is the Borel transform of the quantity
$F(s) = - (9/(4 \pi^2)) E_{b \bar b}(s)/({\overline m}_b a(\mu_s))$
[in analogy with $S$ of (\ref{mbexp})]. Since now the TPS of
$R_{F(s)}$ is known to $\sim\!b^3$, the Pad\'e ${\rm P[2/1]}(b)$
thereof can be taken; using then the value (\ref{Nmnf4})
of $N_m$, the renormalon cancellation condition (\ref{NmEqq})
gives numerically the $s$-$us$ separation parameter
\begin{equation}
\kappa = 0.59 \pm 0.19 \ .
\label{kappa}
\end{equation}
It was possible to obtain the value of $\mu_f$ ($\Leftrightarrow \kappa$)
because the dependence on $\mu_f$ in $N_m$ of Eq.~(\ref{NmEqq}) was taken
(and is known) only to the leading order, the ultrasoft part
was excluded, and $N_m$ is well-known (\ref{Nmnf4}). 
This is similar to the scale-fixing in the
effective charge (ECH) method \cite{ECH}.
If the ultrasoft contributions are included in Eq.~(\ref{NmEqq}),
the value (\ref{Nmnf4}) of $N_m$ cannot be reproduced. 

\section{Evaluation of the soft contributions}
\label{sec:Esoft}

Knowing now the expansion of 
$F(s) = - (9/(4 \pi^2)) E_{b \bar b}(s)/({\overline m}_b a(\mu_s))$
up to $\sim\!a^4$, the Borel transform of this
quantity can be constructed, e.g., with the approach of
the ``$\sigma$-regularized'' bilocal expansion \cite{CCG},
which is a generalization of the bilocal expansion (\ref{Bmb2})
\begin{eqnarray}
\lefteqn{
B_{F(s)}(b) = 
\frac{ 9 N_m  \mu_s}{2 \pi {\overline m}_b a(\mu_s) 
(1\!-\!2 b)^{1\!+\!\nu}}\!
\left[ \sum_{k=0}^{\infty} {\tilde {\cal C}}_k (1\!-\!2 b)^k \right] }
\nonumber\\
&& \times \exp\left[ - \frac{1}{8 \sigma^2} (1\!-\!2 b)^2 \right] +
B_{F(s)}^{\rm (an.)}(b) \ .
\label{Bfs}
\end{eqnarray}
The exponential was introduced in order to suppress the
renormalon part away from $b \approx 1/2$.
The first four coefficients ${\tilde {\cal C}}_k$ are known
(${\tilde {\cal C}}_0=1$), and the analytic part is known now up to
$\sim\!b^3$. The analytic part we can
evaluate either as TPS or as Pad\'e ${\rm P[2/1]}(b)$.
The requirement of the absence of the pole around
$b\!=\!1/2$ in that part, and the independence (weak dependence)
on the renormalization scale $\mu_s$ for the 
Borel-resummed result $E_{b \bar b}(s)$, lead us to
fix the $\sigma$ parameter to the values $\sigma = 0.36 \pm 0.03$.
The Borel integration is performed as in
Eq.~(\ref{BSint}), with the ray (PV) path prescription taken. 

\section{Estimate of the ultrasoft contribution}
\label{sec:Eus}
The ultrasoft part of the energy is known
only to the leading order ($\sim\!m_b a^5$)
\begin{eqnarray}
E_{b \bar b}(us)^{\rm (p)} &\approx& 
- \frac{4}{9} {\overline m}_q \pi^2 f_{3}(us) a^5(\mu_{us})
\nonumber\\
&\approx & (- 150 \pm 100) \ {\rm MeV} \ .
\label{Ebbusp}
\end{eqnarray}
Here, $f_3(us; \mu_f)$ was determined in Sec.~\ref{sec:sep};
the ultrasoft renormalization scale $\mu_{us}$ should be
$\sim\!\alpha_s^2 m_b$, but was taken
numerically to be higher, in the soft regime
($\mu \approx 1.5$-$2.0$ GeV $\ \Rightarrow
\ \alpha_s(\mu) \approx 0.30$-$0.35$), because perturbative
QCD does not allow a running to very low scales.
The bottom mass value was taken ${\overline m}_b = 4.2$ GeV.
The nonperturbative contribution comes primarily from the
gluonic condensate and gives $E_{b \bar b}(us)^{\rm (np)}
\approx 50 \pm 35$ MeV if the gluon condensate values
$\langle (\alpha_s/\pi) G^2 \rangle =
0.009 \pm 0.007 \ {\rm GeV}^4$  \cite{Ioffe:2002be}
are taken. This then results in the following estimate
of the ultrasoft contributions to the binding energy
\begin{equation}
E_{b \bar b}(us)^{\rm (p+np)} \approx (-100 \pm 106) \ {\rm MeV} \ .
\label{Ebbus}
\end{equation} 
In addition, there are contributions to the 
$\Upsilon(1S)$ mass 
due to the nonzero mass of the charm quark
\cite{Brambilla:2001qk}
$\delta M_{\Upsilon}(1S, m_c \not=0) \approx 25 \pm 10$ MeV.

\section{Extraction of the mass ${\overline m}_b$}
\label{sec:concl}

Adding together the Borel-resummed values $2 m_b$,
$E_{b \bar b}(s)$ and $E_{b \bar b}(us)$,
requiring the reproduction of the measured mass
value $M_{\Upsilon}(1S)$ (with the mentioned
$m_c\!\not=\!0$ effect subtracted), 
we extract the following value
for the mass ${\overline m}_b \equiv
{\overline m}_b(\mu\!=\!{\overline m}_b)$:
\begin{equation}
{\overline m}_b({\overline m}_b) = 4.241 \pm 0.068 \ {\rm GeV} \ ,
\label{mbbav2}
\end{equation}
when the QCD coupling value is taken as
$\alpha_s(M_Z) = 0.1180 \pm 0.0015$.
The major source of uncertainty in the result
(\ref{mbbav2}) is the uncertainty from the
ultrasoft contributions (\ref{Ebbus})
($\pm 0.049$ GeV). The other appreciable uncertainties
are from the ambiguity of the soft renormalization
scale $\mu_s\!=\!3\!\pm\!1$ GeV ($\pm 0.013$ GeV) and
of $\alpha_s(M_Z) = 0.1180 \pm 0.0015$
($\pm 0.013$ GeV). If the central value of the gluon
condensate $\langle (\alpha_s/\pi) G^2 \rangle$
is increased from $0.009$ \cite{Ioffe:2002be} to
$0.024 \ {\rm GeV}^4$ 
(used in \cite{Corcella:2002uu,Eidemuller:2002wk}), 
the central value (\ref{mbbav2}) decreases
to ${\overline m}_b({\overline m}_b) = 4.204$ GeV.
This is close to the values of QCD spectral sum rule
calculations which gave central values
${\overline m}_b({\overline m}_b) = 4.20$ GeV
\cite{Corcella:2002uu}; $4.24$ GeV \cite{Eidemuller:2002wk};
and ${\overline m}_b(m_b)=4.23$ GeV \cite{Narison:1994ag}
(Ref.~\cite{Narison:1994ag} uses central condensate value
$\langle (\alpha_s/\pi) G^2 \rangle =0.019$).
The TPS evaluation of $M_{\Upsilon}(1S)$,
without accounting for the renormalon
problem, extracts higher central values 
${\overline m}_b({\overline m}_b) = 4.349$ 
\cite{Penin:2002zv}.

\end{document}